\documentstyle[fleqn,run2col,epsfig]{article}

\newcommand{\AmS}{{\protect\the\textfont2
  A\kern-.1667em\lower.5ex\hbox{M}\kern-.125emS}}

\title{Constraints on the Gluon Density from Lepton Pair Production}

\author{E.~L.~Berger\address{HEP Theory Group, Argonne National
        Laboratory, 9700 South Cass Avenue, Argonne, IL 60439,
        USA}%
        \thanks{Supported by the U.S.\ Department of Energy,
        Division of High Energy Physics, under Contract
        W-31-109-ENG-38.}
        and 
        M.~Klasen\address{II.\ Insitut f\"ur Theoretische
        Physik, Universit\"at Hamburg, Luruper Chaussee 149, D-22761
        Hamburg, Germany}%
        \thanks{Supported by Bundesministerium f\"ur Bildung
        und Forschung under Contract 05 HT9GUA 3, by Deutsche
        Forschungsgemeinschaft under Contract KL 1266/1-1, and by the
        European Commission under Contract ERBFMRXCT980194.}}
       
\begin{document}

\begin{abstract}
The hadroproduction of lepton pairs with mass $Q$ and finite transverse
momentum $Q_T$ is described in perturbative QCD by the same partonic
subprocesses as prompt photon production. We demonstrate that, like prompt
photon production, lepton pair production is dominated by quark-gluon
scattering in the region $Q_T>Q/2$. This feature leads to sensitivity to the 
gluon density in kinematical regimes accessible in collider and 
fixed target experiments, and it provides a new independent method for 
constraining the gluon density.
\end{abstract}

\maketitle

% mkl preprint number
\vspace*{-7.2cm} \noindent ANL-HEP-CP-00-002 \\
\noindent DESY 00-006
\vspace*{ 5.7cm}

\section{INTRODUCTION}
The production of lepton pairs in hadron collisions $h_1h_2\rightarrow\gamma^*
X;\gamma^*\rightarrow l\bar{l}$ proceeds through an intermediate
virtual photon via $q {\bar q} \rightarrow \gamma^*$, and the subsequent 
leptonic decay of the virtual photon. Traditionally, interest in
this Drell-Yan process has concentrated on lepton pairs with
large mass $Q$ which justifies the application of perturbative QCD and allows 
for the extraction of the antiquark density in hadrons \cite{Drell:1970wh}.

Prompt photon production $h_1h_2\rightarrow\gamma X$ can be calculated in
perturbative QCD if the transverse momentum $Q_T$ of the photon is
sufficiently large. Because the quark-gluon Compton subprocess is dominant, 
$g q \rightarrow \gamma X$, this reaction provides essential information on the
gluon density in the proton at large $x$ \cite{Martin:1998sq}. Unfortunately,
the analysis suffers from fragmentation, isolation, and intrinsic transverse
momentum uncertainties. Alternatively, the gluon density can be constrained
from the production of jets with large transverse momentum at hadron colliders
\cite{Lai:1999wy}, but the information from different experiments and colliders 
is ambiguous.

In this paper we demonstrate that, like prompt photon production, lepton pair
production is dominated by quark-gluon scattering in the region $Q_T>Q/2$.
This realization means that new independent constraints on the gluon density 
may be derived from Drell-Yan data in kinematical regimes that are accessible 
in collider and fixed target experiments but without the theoretical and 
experimental uncertainties present in the prompt photon case.

In Sec.~\ref{sec:2}, we review the relationship between virtual and
real photon production in hadron collisions in next-to-leading order QCD.
In Sec.~\ref{sec:3} we present our numerical results, and Sec.~\ref{sec:4}
is a summary.

\section{NEXT-TO-LEADING ORDER QCD FORMALISM}
\label{sec:2}

In leading order (LO) QCD, two partonic subprocesses contribute to the
production of virtual and real photons with non-zero transverse momentum:
$q\bar{q}\rightarrow\gamma^{(*)}g$ and $qg\rightarrow\gamma^{(*)}q$.
The cross section for lepton pair production is related to the cross section
for virtual photon production through the leptonic branching ratio of the
virtual photon $\alpha/(3\pi Q^2)$. The virtual photon cross section reduces
to the real photon cross section in the limit $Q^2\rightarrow 0$.

The next-to-leading order (NLO) QCD corrections arise from virtual one-loop
diagrams interfering with the LO diagrams and from real emission diagrams. At
this order $2 \rightarrow 3$ partonic processes with incident gluon pairs $(gg)$, 
quark pairs $(qq)$, and non-factorizable quark-antiquark $(q\bar{q}_2)$ processes 
contribute also.  Singular contributions are regulated in $n$=4-2$\epsilon$ 
dimensions and removed through $\overline{\rm MS}$ renormalization, factorization, 
or cancellation between virtual and real contributions. An important difference
between virtual and real photon production arises when a quark emits a
collinear photon. Whereas the collinear emission of a real photon leads to a
$1/\epsilon$ singularity that has to be factored into a fragmentation
function, the collinear emission of a virtual photon yields a finite
logarithmic contribution since it is regulated naturally by the photon
virtuality $Q$. In the limit $Q^2\rightarrow 0$ the NLO virtual photon
cross section reduces to the real photon cross section if this logarithm is
replaced by a $1/\epsilon$ pole. A more detailed discussion can be found
in \cite{Berger:1998ev}.

The situation is completely analogous to hard
photoproduction where the photon participates in the scattering in the initial
state instead of the final state. For real photons, one encounters an
initial-state singularity that is factored into a photon structure function.
For virtual photons, this singularity is replaced by a logarithmic dependence
on the photon virtuality $Q$ \cite{Klasen:1998jm}.

A remark is in order concerning the interval in $Q_T$ in which our analysis is 
appropriate.  In general, in two-scale situations, a series of logarithmic 
contributions will arise with terms of the type $\alpha_s^n \ln^n (Q/Q_T)$.  Thus, 
if either $Q_T >> Q$ or $Q_T << Q$, resummations of this series must be considered. 
For practical reasons, such as event rate, we do not venture into the domain 
$Q_T >> Q$, and our fixed-order calculation should be adequate.  On the 
other hand, the cross section is large in the region $Q_T << Q$.  In previous 
papers~\cite{Berger:1998ev}, we compared our cross sections with available 
fixed-target and collider data on massive lepton-pair production, and we were able
to establish that fixed-order perturbative calculations, without resummation, 
should be reliable for $Q_T > Q/2$.  At smaller values of $Q_T$, non-perturbative 
and matching complications introduce some level of phenomenological ambiguity.  For 
the goal we have in mind, viz., contraints on the gluon density, it would appear 
best to restrict attention to the region $Q_T \geq Q/2$, but below $Q_T >> Q$.

\section{PREDICTED CROSS SECTIONS}
\label{sec:3}

In this section we present numerical results for the production of lepton pairs
in $p\bar{p}$ collisions at the Tevatron with center-of mass energy
$\sqrt{S}=1.8$ and 2.0 TeV.
We analyze the invariant cross section $Ed^3\sigma/
dp^3$ averaged over the rapidity interval -1.0 $<y<$ 1.0.
We integrate the cross section over various intervals of pair-mass 
$Q$ and plot it as a function of the transverse momentum $Q_T$.
Our predictions are based on a NLO QCD calculation \cite{Arnold:1991yk} and
are evaluated in the $\overline{\rm MS}$ renormalization scheme. The
renormalization and factorization scales are set to $\mu=\mu_f=
\sqrt{Q^2+Q_T^2}$. If not stated otherwise, we use the CTEQ4M
parton distributions \cite{Lai:1997mg} and the corresponding value of
$\Lambda$ in the two-loop expression of $\alpha_s$ with four flavors (five if
$\mu>m_b$). The Drell-Yan factor $\alpha/(3\pi Q^2)$ for the decay of the
virtual photon into a lepton pair is included in all numerical results.

\begin{figure}[htb]
 \begin{center}
  {\unitlength1cm
  \begin{picture}(7.6,10.5)
   \epsfig{file=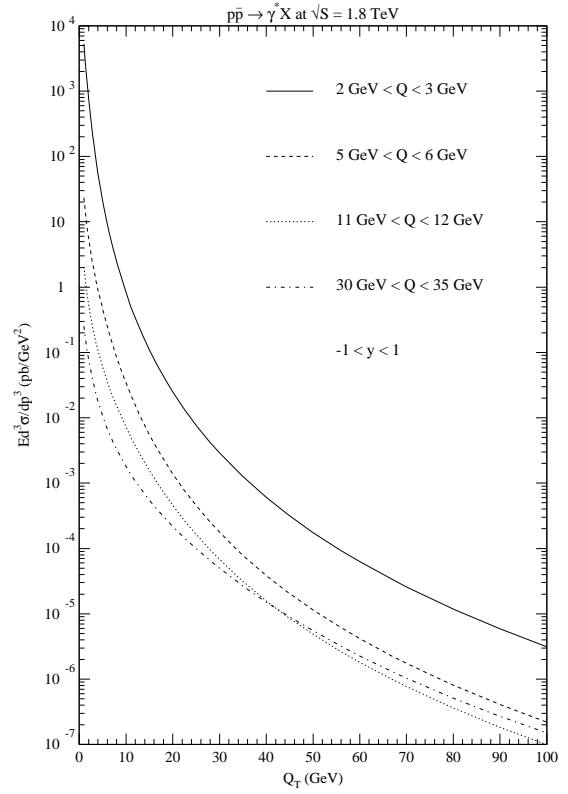,bbllx=60pt,bblly=100pt,bburx=495pt,bbury=725pt,%
           height=10.5cm}
  \end{picture}}
 \end{center}
\vspace*{-1cm}
\caption{Invariant cross section $Ed^3\sigma/dp^3$ as a function of $Q_T$
for $p\bar{p} \rightarrow \gamma^* X$ at $\sqrt{S}=1.8$ TeV in
non-resonance regions of $Q$. The cross section falls with the mass of the
lepton pair $Q$ and, more steeply, with its transverse momentum $Q_T$.}
\label{fig:1}
\end{figure}

In Fig.~\ref{fig:1} we display the NLO QCD cross section for lepton pair
production at the Tevatron at $\sqrt{S}=1.8$ TeV as a function of $Q_T$ for
four regions of $Q$. The regions of $Q$ have been chosen to avoid
resonances, {\it i.e.\ } between $2$ GeV and the $J/\psi$ resonance,
between the $J/\psi$ and the $\Upsilon$ resonances, above the $\Upsilon$'s,
and a high mass region. The cross section falls both with the mass of the
lepton pair $Q$ and, more steeply, with its transverse momentum $Q_T$.
No data are available yet from the CDF and D0 experiments.  However, prompt 
photon production data exist to $Q_T\simeq 100$ GeV, where the cross section
is about $10^{-3}$ pb/GeV$^2$. It should be possible to analyze Run I
data for lepton pair production to at least $Q_T\simeq 30$ GeV where one
can probe the parton densities in the proton up to $x_T = 2Q_T/\sqrt{S}\simeq
0.03$. The UA1
collaboration measured the transverse momentum distribution of lepton
pairs at $\sqrt{S}=630$ GeV up to $x_T=0.13$ \cite{Albajar:1988iq}, and their
data agree well with our theoretical results \cite{Berger:1998ev}.

\begin{figure}[htb]
 \begin{center}
  {\unitlength1cm
  \begin{picture}(7.6,10.5)
   \epsfig{file=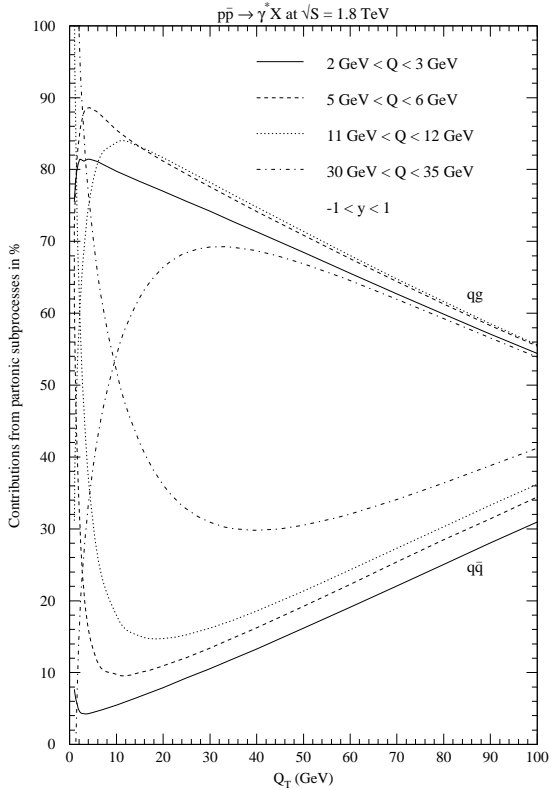,bbllx=60pt,bblly=100pt,bburx=495pt,bbury=725pt,%
           height=10.5cm}
  \end{picture}}
 \end{center}
\vspace*{-1cm}
\caption{Contributions from the partonic subprocesses $qg$ and $q\bar{q}$ to
the invariant cross section $Ed^3\sigma/dp^3$ as a function of $Q_T$
for $p\bar{p}\rightarrow \gamma^* X$ at $\sqrt{S}$ = 1.8 TeV. The
$qg$ channel dominates in the region $Q_T > Q/2$.}
\label{fig:2}
\end{figure}

The fractional contributions from the $qg$ and $q\bar{q}$ subprocesses 
through NLO are shown in Fig.~\ref{fig:2}. It is evident 
that the $qg$ subprocess is the most important subprocess as long as
$Q_T > Q/2$. The dominance of the $qg$ subprocess diminishes somewhat with $Q$,
dropping from over 80 \% for the lowest values of $Q$ to about 70 \%
at its maximum for $Q \simeq$ 30 GeV. In addition, for very large $Q_T$, the
significant luminosity associated with the valence dominated $\bar{q}$
density in $p\bar{p}$ reactions begins to raise the fraction of the cross
section attributed to the $q\bar{q}$ subprocesses.
Subprocesses other than those initiated by the $q\bar{q}$ and
$q g$ initial channels are of negligible import.

\begin{figure}[htb]
 \begin{center}
  {\unitlength1cm
  \begin{picture}(7.6,10.5)
   \epsfig{file=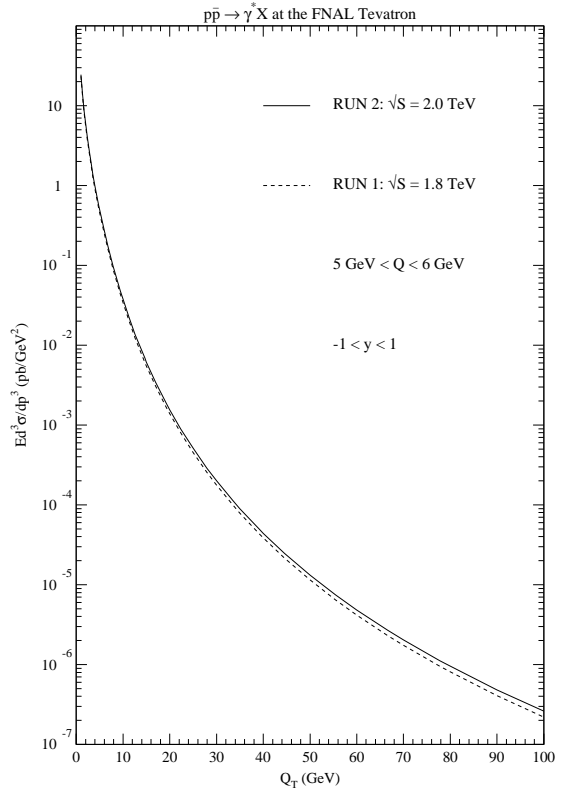,bbllx=60pt,bblly=100pt,bburx=495pt,bbury=725pt,%
           height=10.5cm}
  \end{picture}}
 \end{center}
\vspace*{-1cm}
\caption{Invariant cross section $Ed^3\sigma/dp^3$ as a function of
$Q_T$ for $p\bar{p} \rightarrow \gamma^* X$ and two different
center-of-mass energies of the Tevatron (Run 1: $\sqrt{S}=1.8$ TeV,
Run 2: $\sqrt{S}=2.0$ TeV). The cross section for Run 2 is
5 to 20 \% larger, depending on $Q_T$.}
\label{fig:7}
\end{figure}

We update the Tevatron center-of-mass energy to Run II conditions 
($\sqrt{S}= 2.0$ TeV) and use the latest global fit by the CTEQ 
collaboration (5M). Figure~\ref{fig:7}
demonstrates that the larger center-of-mass energy increases the
invariant cross section
for the production of lepton pairs with mass 5 GeV $<Q<$ 6 GeV by
5 \% at low $Q_T \simeq 1$ GeV and 20 \% at high $Q_T \simeq 100$ GeV. 
In addition, the expected luminosity for Run II of 2 fb$^{-1}$ should
make the cross section accessible to $Q_T\simeq 100$ GeV or
$x_T\simeq 0.1$. This extension would constrain the gluon density in the 
same regions as prompt photon production in Run I.

\setcounter{footnote}{0}

Next we present a study
of the sensitivity of collider and fixed target experiments to the gluon
density in the proton. The full uncertainty in the gluon density is not known.
Here we estimate this uncertainty from
the variation of different recent parametrizations. We choose the latest
global fit by the CTEQ collaboration (5M) as our point of reference
\cite{Lai:1999wy} and compare results to those based on their preceding analysis 
(4M\cite{Lai:1997mg}) and on a fit with a higher gluon density (5HJ) intended to
describe the CDF and D0 jet data at large transverse momentum.  We also compare 
to results based on global fits by MRST \cite{Martin:1998sq}, who provide three 
different sets with a central, higher, and lower gluon density, and to GRV98
\cite{Gluck:1998xa}\footnote{In this set a purely perturbative generation of
heavy flavors (charm and bottom) is assumed. Since we are working in a massless
approach, we resort to the GRV92 parametrization for the charm contribution
\cite{Gluck:1992ng} and assume the bottom contribution to be negligible.}.

\begin{figure}[htb]
 \begin{center}
  {\unitlength1cm
  \begin{picture}(7.6,10.5)
   \epsfig{file=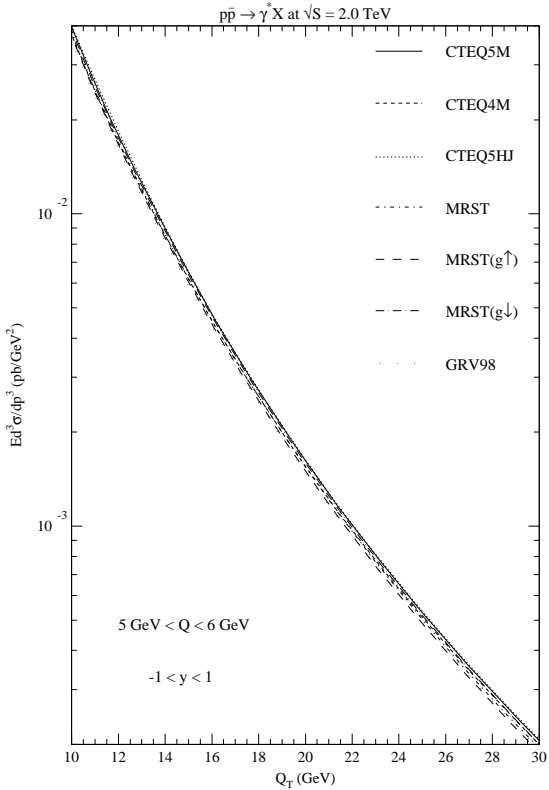,bbllx=60pt,bblly=100pt,bburx=495pt,bbury=725pt,%
           height=10.5cm}
  \end{picture}}
 \end{center}
\vspace*{-1cm}
\caption{Invariant cross section $Ed^3\sigma/dp^3$ as a function of $Q_T$
for $p\bar{p} \rightarrow \gamma^* X$ at $\sqrt{S}=2.0$ TeV in the
region between the $J/\psi$ and $\Upsilon$ resonances. The largest differences
from CTEQ5M are obtained with GRV98 at low $Q_T$ (minus 10 \%) and with
MRST(g$\uparrow$) at large $Q_T$ (minus 7 \%).}
\label{fig:5}
\end{figure}

In Fig.~\ref{fig:5} we plot the cross section for lepton pairs with mass 
between the
$J/\psi$ and $\Upsilon$ resonances at Run II of the Tevatron in the region
between $Q_T=10$ and 30 GeV ($x_T = 0.01 \dots 0.03$). For the CTEQ
parametrizations we find that the cross section increases from 4M to 5M by 2.5
\% ($Q_T=30$ GeV) to 5 \% ($Q_T=10$ GeV) and from 5M to 5HJ by 1 \% in the
whole $Q_T$-range. The largest differences from CTEQ5M are obtained with GRV98 at
low $Q_T$ (minus 10 \%) and with MRST(g$\uparrow$) at large $Q_T$ (minus 7\%).

\begin{figure}[htb]
 \begin{center}
  {\unitlength1cm
  \begin{picture}(9,8.08)
   \epsfig{file=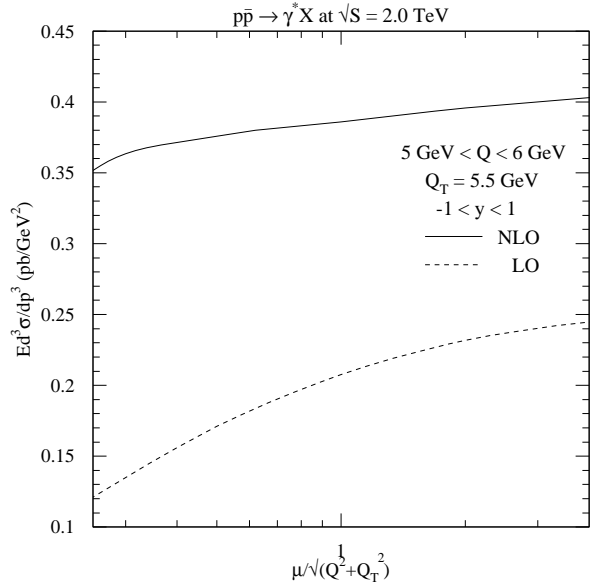,height=9cm}
  \end{picture}}
 \end{center}
\vspace*{-1cm}
\caption{Invariant cross section $Ed^3\sigma/dp^3$ as a function of
the renormalization and factorization scale $\mu=\mu_f$ for $p\bar{p}
\rightarrow \gamma^* X$ at $\sqrt{S}=2.0$ TeV in the region between
the $J/\psi$ and $\Upsilon$ resonances and $Q_T=5.5$ GeV.
In the interval $0.5 <
\mu/\sqrt{Q^2+Q_T^2} < 2$ the dependence of the cross section on the
scale $\mu=\mu_f$ drops from $\pm 15\%$ (LO) to $\pm 2.5\%$ (NLO).
The $K$-Factor (NLO/LO) is approximately 2.}
\label{fig:8}
\end{figure}

The theoretical uncertainty in the cross section can be estimated by varying
the renormalization and factorization scale $\mu=\mu_f$ around the central
value $\sqrt{Q^2+Q_T^2}$. Figure~\ref{fig:8}
shows this variation for $p\bar{p}
\rightarrow \gamma^* X$ at $\sqrt{S}=2.0$ TeV in the region between
the $J/\psi$ and $\Upsilon$ resonances. In the interval $0.5 <
\mu/\sqrt{Q^2+Q_T^2} < 2$ the dependence of the cross section on the
scale $\mu=\mu_f$ drops from $\pm 15\%$ (LO) to the small value 
$\pm 2.5\%$ (NLO). The $K$-factor ratio (NLO/LO) is approximately 2, 
as one might expect naively.  

\begin{figure}[htb]
 \begin{center}
  {\unitlength1cm
  \begin{picture}(7.6,10.5)
   \epsfig{file=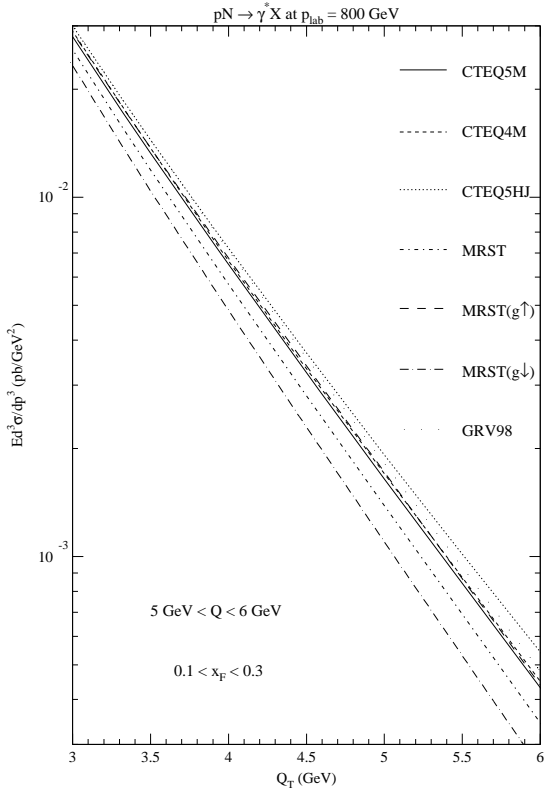,bbllx=60pt,bblly=100pt,bburx=495pt,bbury=725pt,%
           height=10.5cm}
  \end{picture}}
 \end{center}
\vspace*{-1cm}
\caption{Invariant cross section $Ed^3\sigma/d p^3$ as a function of 
$Q_T$ for $p N \rightarrow \gamma^* X$ at $p_{\rm lab}=$ 800 GeV. The cross
section is highly sensitive to the gluon distribution in the proton in regions
of $x_T$ where it is poorly constrained in current analyses.}
\label{fig:6}
\end{figure}

A similar analysis for Fermilab's fixed target experiment E772
\cite{McGaughey:1994dx} is shown in Fig.~\ref{fig:6}.
In this experiment, a deuterium target is bombarded with a proton beam of
momentum $p_{\rm lab}=$ 800 GeV, {\it i.e.} $\sqrt{S}=38.8$ GeV. The cross
section
is averaged over the scaled longitudinal momentum interval 0.1 $< x_F <$ 0.3.
In fixed target experiments one probes substantially larger regions of $x_T$
than in collider experiments.
Therefore one expects greater sensitivity to the gluon distribution in
the proton. We find that use of CTEQ5HJ increases the cross section by 7 \%
(26 \%) w.r.t.\ CTEQ5M at $Q_T=3$ GeV ($Q_T=6$ GeV) and by 134 \% at
$Q_T=10$ GeV. With MRST(g$\downarrow$) the cross section drops relative to 
the CTEQ5M-based values  
by 17 \%, 40 \%, and 59 \% for these three choices of $Q_T$. 

\begin{figure}[htb]
 \begin{center}
  {\unitlength1cm
  \begin{picture}(9,7.68)
   \epsfig{file=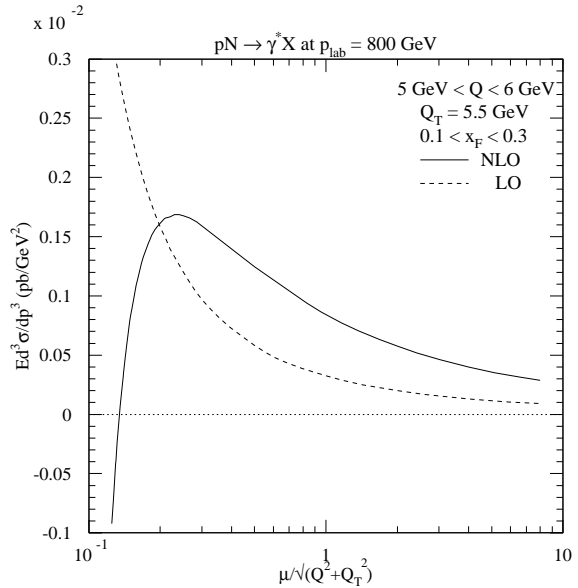,height=8.6cm}
  \end{picture}}
 \end{center}
\vspace*{-1cm}
\caption{Invariant cross section $Ed^3\sigma/d p^3$ as a function of 
the renormalization and factorization scale $\mu=\mu_f$ for $p N \rightarrow
\gamma^* X$ at $p_{\rm lab}=$ 800 GeV. In the interval
$0.5 < \mu/\sqrt{Q^2+Q_T^2} < 2$ the dependence of the cross section
on the scale $\mu$ drops from $\pm 49\%$ (LO) to $\pm 37\%$ (NLO).}
\label{fig:9}
\end{figure}

Figure~\ref{fig:9}
shows the variation of the fixed target cross section on the
renormalization and factorization scale $\mu=\mu_f$. In the interval
$0.5 < \mu/\sqrt{Q^2+Q_T^2} < 2$ the dependence decreases from 
$\pm 49\%$ (LO) to $\pm 37\%$ (NLO).
An optimal scale choice might be $\mu = \mu_f = \sqrt{Q^2+Q_T^2}/4$, where 
the points of Minimal Sensitivity (maximum of NLO) and of Fastest Apparent
Convergence (LO=NLO) nearly coincide.
At $\mu=\mu_f=\sqrt{Q^2+Q_T^2}$, the $K$-factor ratio is 2.6.
The NLO cross section turns negative at the lowest scale shown
$\mu=\mu_f=\sqrt{Q^2+Q_T^2}/8 \simeq 1$ GeV, a value too low to guarantee
perturbative stability.

\section{SUMMARY}
\label{sec:4}
The production of Drell-Yan pairs with
low mass and large transverse momentum is dominated by gluon initiated
subprocesses. In contrast to prompt photon production, uncertainties 
from fragmentation, isolation, and intrinsic transverse momentum are absent.
The hadroproduction of low mass lepton pairs is therefore an advantageous
source of information on the parametrization and size of the gluon density.
The increase in luminosity of Run II increases the
accessible region of $x_T$ from 0.03 to 0.1. The theoretical uncertainty
has been estimated from the scale dependence of the cross sections and
found to be very small for collider experiments. 

\section*{Acknowledgment}
It is a pleasure to thank L.~E.~Gordon for his collaboration.

\end{document}